\documentclass[journal=apchd5,manuscript=article,layout=twocolumn,line=single-spaced]{achemso}

\usepackage{amsmath}
\usepackage{amssymb}
\usepackage{graphicx}
\usepackage{braket}
\usepackage{comment}
\usepackage{hyperref}
\usepackage{soul,color}
\usepackage[bottom]{footmisc}
\usepackage{multicol}

\mciteErrorOnUnknownfalse

\newcommand{\da}{\downarrow}
\newcommand{\ua}{\uparrow}
\newcommand{\ad}{\hat{a}^\dagger}
\newcommand{\an}{\hat{a}^{}}
\newcommand{\bd}{\hat{b}^\dagger}
\newcommand{\bn}{\hat{b}^{}}

\newcommand{\mbf}[1]{\mathbf{#1}}

\newcommand{\EPF}{\ket{\text{EP-F}}}
\newcommand{\EP}{\ket{\text{EP}}}
\newcommand{\PP}{\ket{\text{PP}}}
\newcommand{\tPS}{\ket{\text{2P}}}

\newcommand{\new}[1]{}

\newcommand{\symset}{\mathcal{S}}
\newcommand{\cntset}{\mathcal{P}} 
\newcommand{\indset}{\mathcal{I}} 

\newcommand{\maptoN}{{{\mathcal{M}}}_{N-1 \mapsto N}}
\newcommand{\maptoNa}{{\mathcal{M}}_{N-2 \mapsto N-1}}

\newcommand*{\citen}{}
\DeclareRobustCommand*{\citen}[1]{%
  \begingroup
    \romannumeral-`\x 
    \setcitestyle{numbers}%
    \cite{#1}%
  \endgroup
}

\usepackage{float}

\title{Exact states and spectra of vibrationally dressed polaritons}

\author{M. Ahsan Zeb\new{\footnote{Permanent address: National Centre for Physics, Islamabad 45320, Pakistan}}}
\affiliation{SUPA, School of Physics and Astronomy, University of St Andrews, St Andrews, KY16 9SS, United Kingdom}
\author{Peter G. Kirton}
\affiliation{SUPA, School of Physics and Astronomy, University of St Andrews, St Andrews, KY16 9SS, United Kingdom}
\author{Jonathan Keeling}
\affiliation{SUPA, School of Physics and Astronomy, University of St Andrews, St Andrews, KY16 9SS, United Kingdom}
\email{jmjk@st-andrews.ac.uk}
\date{\today}

\begin{document}

\begin{abstract}
  Strong coupling between light and matter is possible with a variety of
  organic materials. In contrast to the simpler inorganic case, organic
  materials often have a complicated spectrum, with vibrationally dressed
  electronic transitions.    Strong coupling to light competes with this
  vibrational dressing, and if strong enough, can suppress the entanglement
  between electronic and vibrational degrees of freedom.  By exploiting symmetries,  we can perform exact  numerical diagonalization to
  find the polaritonic states for intermediate numbers of
  molecules, and use these to define and validate
  accurate expressions for the lower polariton states and strong-coupling
  spectrum in the thermodynamic limit. Using this approach, we find
  that vibrational decoupling occurs as a sharp transition above
  a critical matter-light coupling strength.   We also demonstrate
  how the polariton spectrum evolves with the number of molecules, recovering
  classical linear optics results only at large $N$.
\end{abstract}

\maketitle

\section*{Introduction}
Strong coupling between matter and light allows one to engineer material
properties to induce desired collective behavior.  A widely studied
example of this is exciton polaritons in semiconductor microcavities.  
Strong coupling between microcavity photons and excitons
produces quasiparticles with ideal properties for Bose condensation at
elevated temperatures~\cite{Kasprzak2006,snoke07science}, 
allowing the study of interacting quantum fluids of light~\cite{Carusotto2013}.
Organic light emitting materials provide a particularly intriguing context
in which to study the effects of  strong coupling.  
Polaritons in such materials are stable
up to room temperature, due to the high binding energy of Frenkel excitons
and the large matter-light coupling strengths~\cite{Agranovich2009a}.  Both
strong coupling~\cite{Lidzey98, Lidzey99,Lidzey00, Tischler05,
  Kena-Cohen08} and polariton condensation or
lasing~\cite{Kena-Cohen10,Plumhof14, Daskalakis14,Michetti15,dietrich16}
have been seen in a wide variety of organic materials, including
J-aggregates, small-molecule crystals, conjugated polymers, and even
biologically produced materials.


Strong matter light coupling with organic materials also enables a range of
further effects\new{~\cite{FlickPNAS17,KowalewskiPNAS17}}, in which material properties of the molecules such as
chemical reaction rate~\cite{Hutchison12,Herrera15,kowalewski16,galego16,galego17}, transport
properties~\cite{orgiu15,Feist15}, 
\new{superradiance~\cite{ZhangCPL17}, photochemistry~\cite{BennettRSC16},}
aspects of the molecular
structure~\cite{Schwartz11,Galego15,Cwik16},
\new{or coupled electron-nuclear dynamics~\cite{FlickJCTC17}}
can be modified by strong
coupling of light to electronic transitions, or by coupling between
infra-red cavities and optically active vibrational
modes~\cite{Shalabney15,Pino15,shalabney15:raman,Pino15a,strashko16}.  Many
of these effects have been associated with the way in which strong matter
light coupling modifies the potential energy landscape for the conformation
of the molecules.  This also gives rise to the complex absorption and
emission spectra observed in organic molecules~\cite{barford13}, which often have large Stokes shifts,
arising from a variety of photophysical effects.  One significant
contribution is the strong vibrational dressing of electronic transitions.
The absorption spectrum is significantly modified in the presence of strong
coupling to light, giving observable signatures of the change of
vibrational conformation.  In all these examples, the behavior observed
results from collective strong coupling, with many molecules coupled to the
same mode of light. Recently strong coupling has also been achieved in the
limit of a small number of molecules, coupled to a plasmonic
resonance~\cite{chikkaraddy16} highlighting the importance of developing
theoretical techniques valid at finite molecule number.

In this paper, we \new{consider the widely used Holstein-Tavis-Cummings model and} give exact results \new{for} how strong matter-light coupling
changes the vibrational configuration of the lower polariton state \new{in this model}.  We
show that there is a sharp crossover in the nature of the polaritonic
ground state as one increases matter-light coupling.
We find 
that
vibrational decoupling 
only occurs above
this critical strength, and so the resulting phenomena, such as
enhanced electron transfer rate as discussed by \citet{Herrera15}, only occur when the polariton splitting is
sufficiently large.  
Most previous work addressing the competition of
strong matter-light coupling and vibrational configuration has either
relied on exact numerics for small numbers of molecules, or used physically
motivated approximations without rigorous derivation.  In contrast,
we present exact numerical results for intermediate numbers of molecules,
and derive analytic approximations which we show closely match the exact 
numerical results.
This means we provide results that can cover the full range of molecule
numbers from $N=1$ to the thermodynamic limit, $N\rightarrow\infty$. 
Indeed, using this approach we show that the polariton spectrum
for small $N$ and large $N$ is quite distinct, and that linear optics
results are only recovered in the large $N$ limit.

\section*{Modeling organic polaritons}

Collective effects due to matter-light interaction have been studied
extensively in the context of superradiance, lasing and polariton
condensation~\cite{chiocchetta17}.  The archetypal models used to study this
are the Tavis-Cummings~\cite{Tavis1968} and
Dicke~\cite{Dicke1954,Garraway2011a} models, which describe many two-level
systems coupled to light.  However organic molecules are not two-level
systems, and as discussed above, 
we must include the coupling between electronic states and molecular
conformation, which produces vibrational sidebands in
the absorption and emission spectra.  
In order to include such physics, we consider the Holstein-Tavis-Cummings (HTC) model
which has been introduced and used in several previous works on strong
coupling with organic
molecules~\cite{cwik14,Spano15,Galego15,Cwik16,Herrera15,wu2016}.  
Such a model allows one to find how strong matter-light coupling
interacts with the formation of polarons --- i.e.\ dressing of
electronic states by vibrational excitations.


\begin{figure}[htpb]
  \centering
  \includegraphics[width=3.2in]{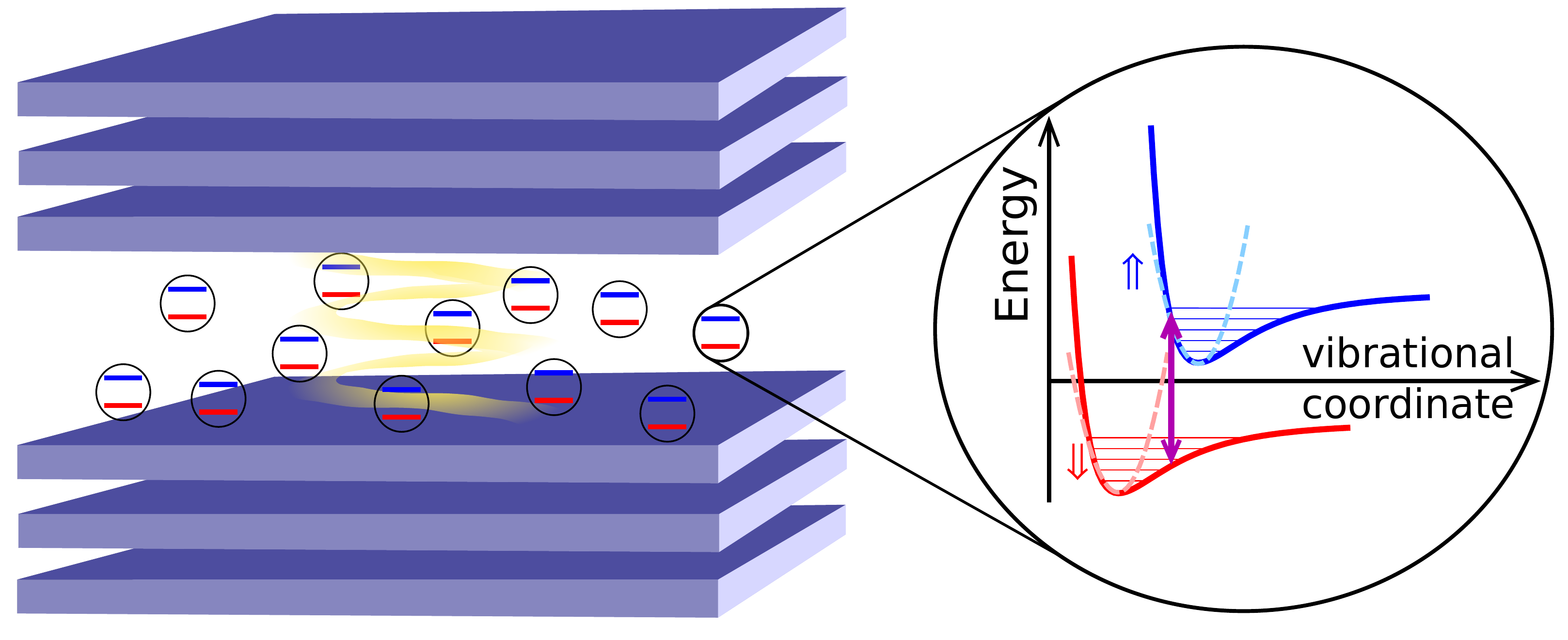}
  \caption{Cartoon of the Holstein-Tavis-Cummings model, Eq.~\eqref{eq:dh}.  We consider
    $N$ molecules placed in a cavity, where each molecule is represented by
    two electronic states, and an associated vibrational mode corresponding
    to the potential energy surface in the given electronic state.}
  \label{fig:htc-cartoon}
\end{figure}

The HTC model, illustrated in Fig.~\ref{fig:htc-cartoon} is given by:
\begin{multline}
  \label{eq:dh}
  \hat H
  =
  \omega \ad \an
  +
  \sum_{i=1}^N \left[
    \omega_0 \hat \sigma^+_i \hat \sigma^-_i
    +
    \frac{\omega_R}{\sqrt{N}} 
    \left( \hat \sigma^+_i \hat a^{} + \hat \sigma^-_i \hat a^\dagger \right)\right.
    \\+\left.
    \omega_v \left(
      \hat b^\dagger_i \hat b^{}_i
      - 
      \lambda_0 \sigma^+_i \hat \sigma^-_i 
      ( \hat b^\dagger_i + \hat b^{}_i ) \right)
  \right].
\end{multline}
The cavity photon mode (annihilation operator $\an$) is coupled to many
molecules, labeled $i=1\ldots N$.  Each molecule is described by two
electronic states (Pauli operator $\hat \sigma_i$), and a (harmonic)
vibrational degree of freedom (annihilation operator $\bn_i$).  \new{The
final term in the Hamiltonian reflects the offset between the optimal vibrational coordinate in the electronic ground state and in the excited electronic state.  Specifically, the dimensionless parameter $\lambda_0$ measures the vibrational coordinate offset in units of the  vibrational harmonic oscillator length (i.e. the size of the ground state vibrational wavefunction).  The parameter $\lambda_0$ is related to the Huang-Rhys parameter $S$ via $S=\lambda_0^2$, which can be understood as the typical number of vibrational quanta excited in an electronic transition of the molecule (in the absence of strong coupling to light).  The parameter $\lambda_0$ can also be related to the Stokes shift between the peak of the absorption spectrum and emission spectrum (in the absence of strong coupling to light).  Roughly, the Stokes shift scales as $\omega_v \lambda_0^2$.}
The matter-light coupling \new{(i.e. coupling between photons and \emph{electronic} transitions of the molecules)} is parameterized by
the collective Rabi splitting $\omega_R$.  

We work in the rotating wave
approximation \new{for this matter-light coupling, hence the number of 
electronic excitations plus photons,}  $N_{\text{ex}} = \ad \an +
\sum_{i=1}^N \hat \sigma^+_i \hat \sigma^-_i$ is conserved. \new{We note that this
approximation requires $\omega_R$ to be small with respect to the
bare cavity and electronic transition frequencies, $\omega$ and $\omega_0$.
Such an approximation is reasonable for many experiments, however there can
exist experiments in the ultra-strong coupling regime where this approximation fails, and counter-rotating terms should be included.  Since our focus in this paper is on the competition of matter-light coupling and the vibrational coupling, we focus on the regime where the rotating wave approximation is valid, as this allows direct comparison to previous results\cite{Herrera15,wu2016}.  While the RWA assumes the coupling $\omega_R$ is weak with respect to $\omega,\,\omega_0$, it does not place any restrictions on the size of the matter-light coupling as compared to the vibrational energy scales $\omega_v$ or $\omega_v \lambda_0^2$.}

The vibrational dressing described by the HTC model is not unique to the
organic molecular systems discussed here.  Similar physics
exists for color centers in diamond~\cite{Doherty2013}, and is also related to models studied in the context of optomechanics, where
systems such as superconducting qubits act
as a transducer between vibrational and optical modes~\cite{pirkkalainen2015, Heikkila2015}.  This connection
between vibrationally dressed electronic transitions in molecules and
optomechanics has been discussed by~\citet{Roelli16}  The methods we
present here can thus form the basis to answering other questions involving
these classes of system (e.g.\ understanding the role of vibronic replicas
of polaritons in condensation and lasing or finding polariton-polariton
interactions), and our analytic ansatz gives new insights into the state of
the $N$ molecule system.

\section*{Results and discussion}

\subsection*{Nature of lower polariton state}

In this section, we present the exact lower polariton state, and discuss
its vibrational configuration.
If one considers a single molecule, i.e.\ $N=1$ in Eq.~(\ref{eq:dh}), the
physics of this model is well understood.  Each sector of fixed $N_{\text{ex}}$
can be mapped to the Rabi model, describing a spin coupled to a bosonic
mode~\cite{irish05,irish07,bera14}.  Our aim is to extend this
understanding to large $N$. In what follows we take two approaches: 
{\it (i)} we describe how, even for relatively large values of $N$, it is possible to
exactly diagonalize the Hamiltonian or time-evolve a state, and    
{\it (ii)} we
compare these exact solutions to analytic approaches.  The exact numerical
results inform the analytic approaches in two ways:  Firstly, studying the
Wigner function of the exact solution, we can identify the nature of
the eigenstates, and construct appropriate variational wavefunctions.  Secondly,
comparison of the exact and variational energies provides confirmation
whether the analytic description is sufficiently accurate.

\begin{figure}[ht]
\includegraphics[width=\columnwidth]{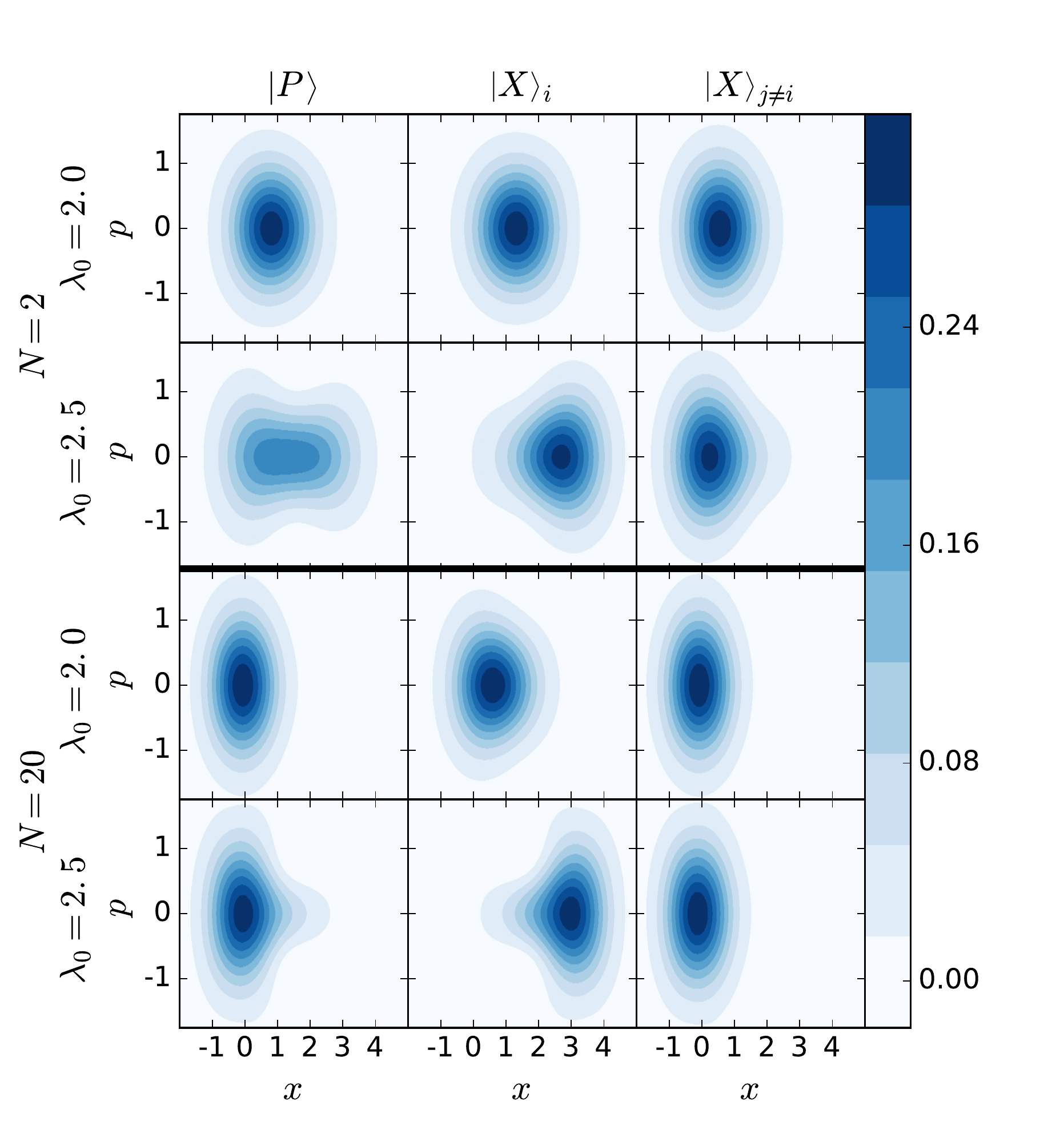}
\caption{Conditional Wigner functions, 
  \new{$W_{\ket{\psi}}(x,p)$} for $N=2$ and $N=20$ at
  $\omega=\omega_0=\omega_R=1$, $\omega_v=0.2$ and $\lambda_0$ as labeled.
  The first column is for a photon present in the cavity 
  \new{$\ket{\psi}=\ket{P}$}, the second
  for the  molecule $i$ being electronically excited ($\ket{X}_i$), and
  the third for molecule $j \neq i$ being excited ($\ket{X}_{j\neq
    i}$).\label{fig:wf}}
\end{figure}

In order to identify the lower polariton state, we find the lowest
energy eigenstate in the one excitation sector, $N_{\text{ex}}=1$ and
examine its vibrational configuration.  We do this in 
Fig.~\ref{fig:wf} by showing the
Wigner function (WF), \new{$W(x,p)$} of the vibrational state of molecule $i$,
conditioned on one of three possible scenarios: a photon is present
(labeled $\ket{P}$), the molecule $i$ is electronically excited
($\ket{X}_i$), or another molecule, $j$ is excited ($\ket{X}_{j\neq i})$.
\new{In plotting these figures, we normalise each Wigner distribution independently.  The ``total'' Wigner function for a given molecular coordinate thus corresponds to a weighted sum of these conditional Wigner functions,
\begin{multline}
  \label{eq:composition}
  W(x,p) = \cos^2(\theta) W_{\ket{P}}(x,p) %
  + \sin^2(\theta) \times\\
  \frac{1}{N}\left[ 
     W_{\ket{X}_i}(x,p) + (N-1) W_{\ket{X}_{j \neq i}}(x,p).
  \right].
\end{multline}
Here the angle $\theta$ parameterizes the exciton--photon composition of the lower polariton state, and the factors of $N$ reflect the fact that the weight in the density matrix of a given molecule being excited is equal and there are $N-1$
molecules $j \neq i$.}
We show
results for both $N=2$ and, a 
large number, $N=20$, and two values of $\lambda_0$, which, as
discussed below are opposite sides of the decoupling crossover.
Details of this exact numerical diagonalization are given in the methods
section below.
 For
$\lambda_0=2.0$ (rows 1 and 3), the vibrational WF is not strongly dependent
on electronic state, and is close to the  vacuum state (i.e.
 a Gaussian centered at $x=0$, $p=0$).  A small displacement is seen for the
$\ket{X}_i$ state at $N=2$, but for $N=20$ this becomes very weak. For such a
state, the vibrational and electronic states factorize, as discussed below.  \new{In the case $N=2$ the small displacement of the peak of the Wigner function away from $x=0$ can be associated to a small admixture of the excited state vibrational configuration.  The suppression of this feature with increasing $N$ is discussed below in the context of the variational ansatz.}
For $\lambda_0=2.5$ (rows 2 and 4), quite distinct
behavior is seen, despite the small change of $\lambda_0$. 
The vibrational and electronic state are
strongly entangled: for $\ket{X}_i$ the vibrational state is very close to a
displaced Gaussian. 
These exact results show that the vibrational decoupling
discussed recently by~\citet{Herrera15} occurs only for small $\lambda_0$.  In that
work it was shown that vibrational decoupling could have dramatic effects
on electron transfer rates.  From results below, we will see that
the crossover value of $\lambda_0$ depends on matter light coupling,
so that the enhancements of energy transfer  can be expected
to ``turn on'' above a critical value of matter-light coupling.

In addition to the vibrational decoupling, Fig.~\ref{fig:wf} for
$N=2, \lambda_0=2.5$ shows another notable feature. We see $\ket{P}$ and $\ket{X}_{j \neq
  i}$ differ qualitatively, that is, even when molecule $i$ is in its electronic
ground state, its vibrational configuration depends on whether other molecules
are excited. For $\ket{P}$ the vibrational
 WF is bimodal, while for $\ket{X}_{j\neq i}$ the vibrational WF
is close to the vacuum state. The bimodal peak can be understood by considering
potential energy surfaces vs vibrational coordinates, 
as discussed by~\citet{Galego15}: for $N=2$, within
the lower polariton manifold the two dimensional vibrational
potential energy surface as a function of $x_i = (\bd_i + \bn_i)/\sqrt{2}$,
has local minima at $(x_1,x_2) = (\sqrt{2}\lambda_0,0),
(0,\sqrt{2}\lambda_0)$. (Note that for $\omega=\omega_0$, there is no minima at
$(0,0)$, so all local minima involve displacements of some coordinate.)  The WF
corresponds to a marginal distribution of the lower polariton wavefunction
after integrating out the other coordinates.  By conditioning on the electron
and photon state, the WF becomes a biased version of this distribution.  For
$\ket{X}_{j \neq i}$, there is a bias toward $x_{j}=\sqrt{2} \lambda_0,
x_{k \neq j} =0$, so the vibrational state for molecule $i \neq j$ is biased
to $0$.  For $\ket{P}$, this bias is absent, and the weights of the two minima
are equal, giving a double peaked structure.  On decreasing $\lambda_0$, or
increasing $\omega_R$, the distinct minima merge~\cite{galego16}, and the
bimodal feature of the vibrational state vanishes replaced by a slightly displaced vacuum.  The
bimodality is not so visible for  $N=20$. 
This is because the $N$ molecule system has
$N$ potential minima, and the
contribution of each minima is suppressed by $1/N$.  We show below that
all of these features can be quantitatively recovered by an appropriate ansatz.

\subsection*{Analytic approximation to lower polariton state}
\label{sec:analyt-appr-ground}

The exact results shown in the previous system reveal how the vibrational
decoupling varies significantly with parameter values, and how for the
entangled state, the wavefunction takes a complicated form.  These results
are however costly to evaluate, and cannot be extended much beyond $N=20$.
In this section, we use insight from these exact results to construct a
variational wavefunction to recover the lower polariton state, and
extrapolate our results from $N \lesssim 20$ to the thermodynamic limit, $N\to \infty$.  We examine in turn a series of
ans\"atze which approximately describe the lower polariton state including
the vibrational mode.  The most crucial feature displayed by
Fig.~\ref{fig:wf} is the entanglement between the electronic and vibrational
states when the coupling to vibrational modes is strong
enough.  To capture this we write the \textit{entangled polaron} ansatz:
\begin{multline}
  \label{eq:EP}
  \EPF =
  \alpha \ket{P} \otimes
  \prod_j \mathcal{D}(\lambda_a,\bn_j)
  \ket{0}_V
  \\
  +
  \frac{\beta}{\sqrt{N}} \sum_i \ket{X}_i \otimes
  \mathcal{D}(\lambda_b,\bn_i)
  \prod_{j \neq i} \mathcal{D}(\lambda_c,\bn_j)
  \ket{0}_V,
\end{multline}
where $\mathcal{D}(\lambda,\bn)=\exp[-\lambda(\bd-\bn)]$ is the vibrational
displacement operator, and $\ket{0}_V$ is the vibrational vacuum state.
This state has distinct vibrational configurations for each of
$\ket{P}$, $\ket{X}_i$ and $\ket{X}_{j\neq i}$ shown in
Fig.~\ref{fig:wf}, allowing the required entanglement between vibrational 
and electronic state.  
The corresponding displacements are given by variational
parameters, $\lambda_a, \lambda_b, \lambda_c$.

\begin{figure}[t]
 \includegraphics[width=\columnwidth]{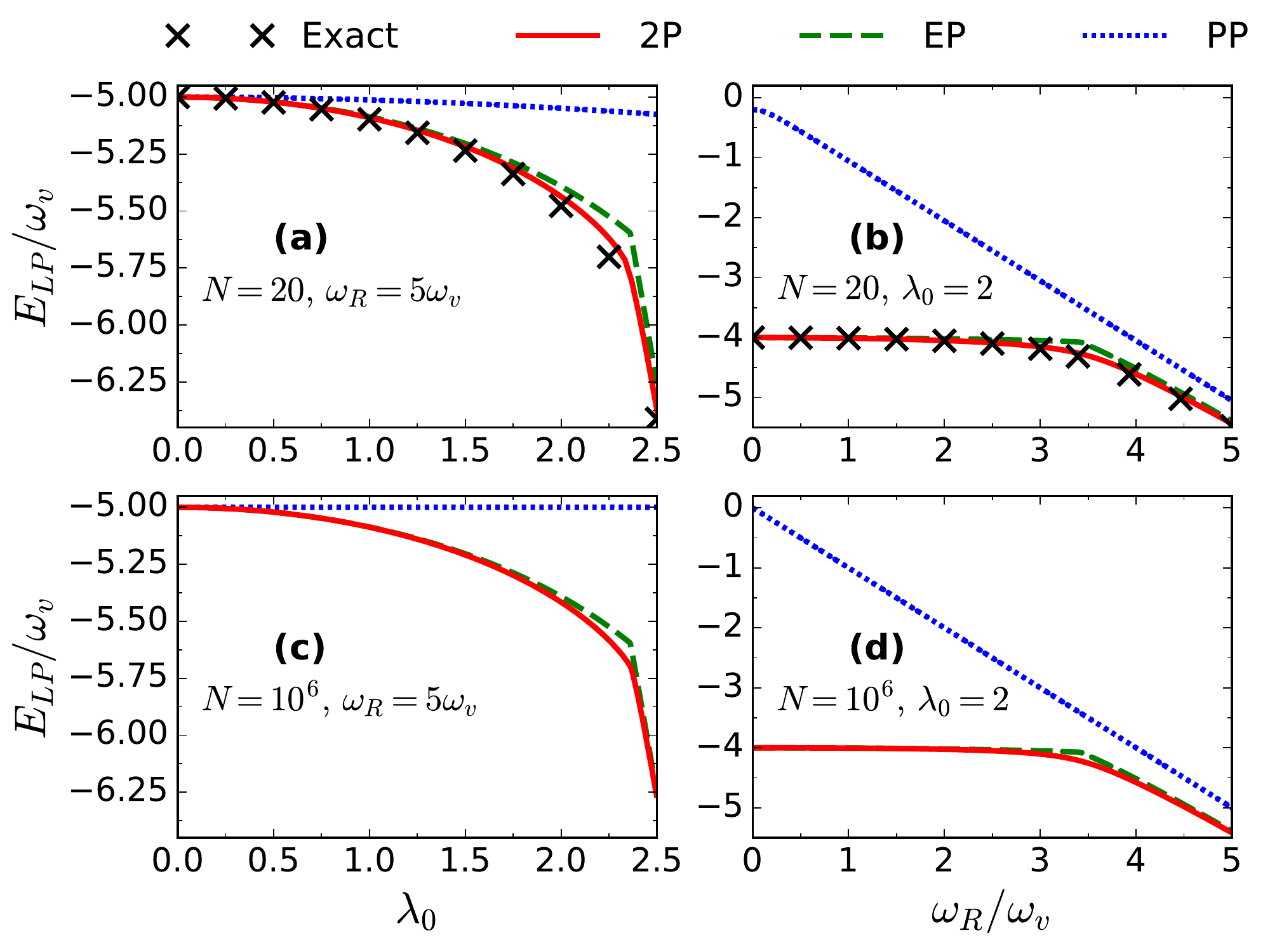}
 \caption{Energy vs $\lambda_0$ and $\omega_R$ for $N=20$ and $N=10^6$. 
   The exact solution is shown as points, while 
    various
   approximate expressions are shown as lines.
   $\omega_0=\omega$ for all panels, and is set as reference for $E_{LP}$.
   \label{fig:ELP}
 }
\end{figure}

As Eq.~(\ref{eq:EP}) is linear in $\alpha,\beta$, one can easily minimize
over these 
 to find:
\begin{equation}
\label{eq:EnEP}
  E_{\text{EP-F}} = 
  \frac{\omega_X + \omega_P}{2}
  - 
  \sqrt{%
    \tilde{\omega}_R^2 +
    \left(  \frac{\omega_X - \omega_P}{2} \right)^2
  },
\end{equation}
with
\begin{equation}
	\tilde{\omega}^2_R = \omega^2_R \exp\left[ - (\lambda_a-\lambda_b)^2 -
  (N-1)(\lambda_a-\lambda_c)^2\right],
\end{equation}
a renormalized Rabi
frequency, and 
\begin{align}
	\omega_X &= \omega_0 + \omega_v(
\lambda_b^2 - 2 \lambda_0\lambda_b + (N-1)\lambda_c^2),\\  
\omega_P
&=\omega+ \omega_v N \lambda_a^2
\end{align}
are renormalized exciton
and photon frequencies.
We will consider two special cases of this ansatz, which we compare to the
exact results in Fig.~\ref{fig:ELP}.  

The first special case is to set
$\lambda_a=\lambda_b=\lambda_c \equiv \lambda$.  This ansatz, which we
refer to as the \textit{product polaron}, $\PP$, matches that
in \citet{Herrera15}.  In this restricted ansatz, one finds that
$\lambda \sim 1/\sqrt{N}$,  and so there is a decoupling between electronic
and vibrational states.  
As seen in Fig.~\ref{fig:ELP}, this ansatz only becomes accurate when the Rabi
frequency is very large compared to $\omega_v \lambda_0^2$ \new{and hence is limited in the range of its applicability.  The $\PP$ ansatz works} in this \new{strong coupling} limit because
one can first diagonalize the matter-light problem, then treat the coupling
to vibrations perturbatively.  The matter-light problem splits into
`bright' polaritonic, and `dark' states.  The bright states are invariant
under permutation of the electronic states of the molecules alone.  In
order to have entanglement between electronic and vibrational states, it is
thus necessary to mix the bright and dark states~\cite{Herrera15},
but when $\omega_R \gg \omega_v \lambda_0^2$, the large bright-dark splitting
prevents vibrational coupling  causing any mixing.

\begin{figure}[ht]
  \includegraphics[width=3.2in]{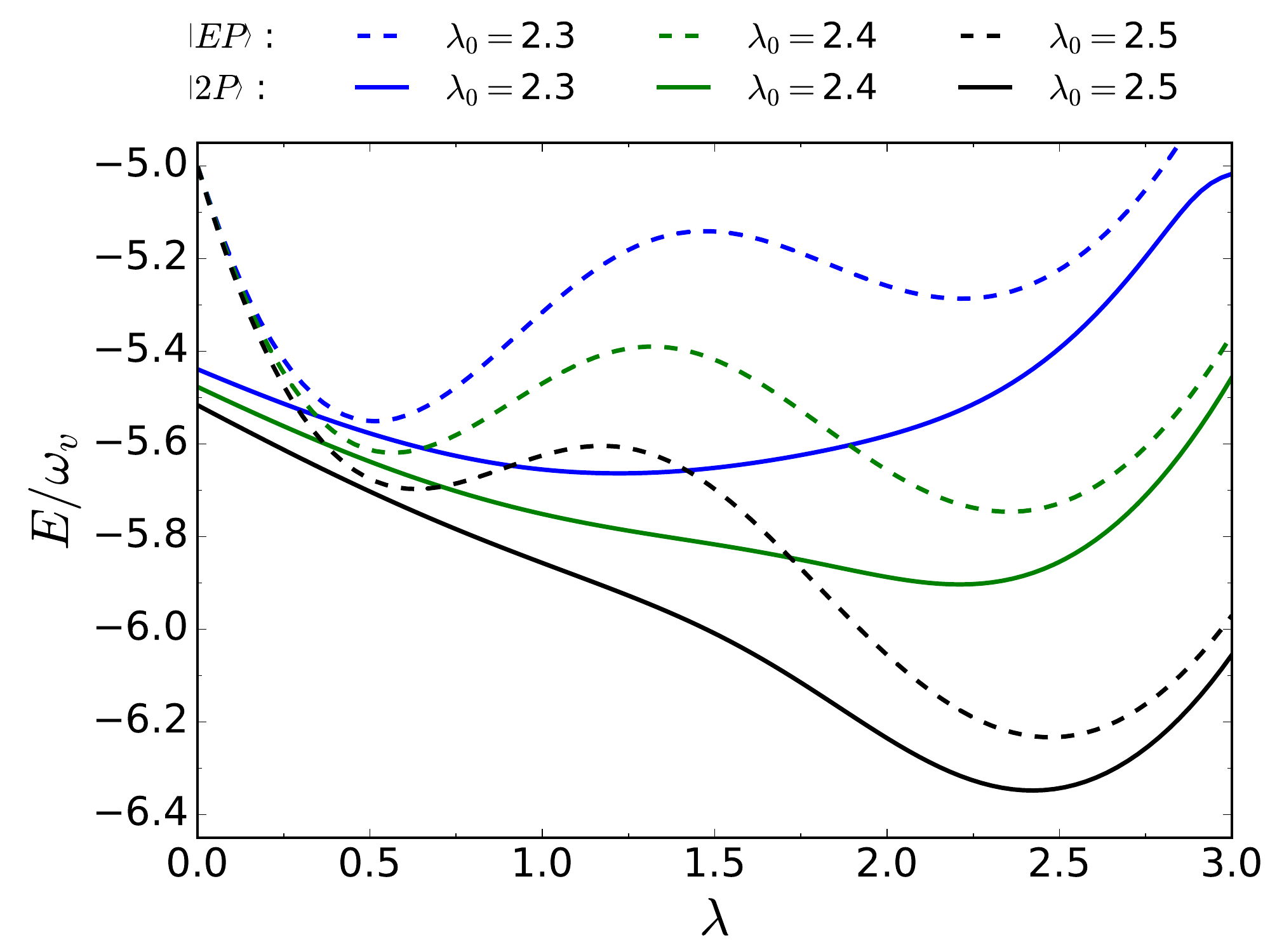}
  \caption{Energy landscape for $\EP$ and $\tPS$ as a function of variational
    parameter $\lambda$, plotted for $N=20$ and $\lambda_0$ as labeled.  For
    $EP$, one sees there is a crossing between the local minima near
    $\lambda_0^2 \simeq \omega_R/\omega_v$, while for $\tPS$ this is smeared
    out.}
  \label{fig:energy}
\end{figure}

An improved restricted ansatz is based on considering the large $N$ limit, making
no assumptions about other parameters. Minimizing the energy clearly requires non-zero
$\tilde{\omega}_R$ and finite $\omega_X, \omega_P$. Hence, we may see that in the
$N\to \infty$ limit one requires $\lambda_a=\lambda_c=0$, but we may keep
non-zero $\lambda_b\equiv\lambda$. This state, $\EP$, leads to an $N$-independent expression for
$E_{\text{EP}}$, which, as can be seen in Fig.~\ref{fig:ELP}, matches the
exact energy well. \new{The form of this ansatz makes clear why any displacement $\lambda_a$ of vibrational coordinates in the photon sector is suppressed at large $N$.} For all parameters, the ansatz $\EP$ is closer to the exact solution than $\PP$.  The only
notable deviation between $\EP$ and the exact solution occurs near
$\omega_R \simeq \omega_v \lambda_0^2$, where there is a kink for $\EP$ due
to a crossing between the
energies of the local minima near $\lambda \simeq 0$ and $\lambda \simeq
\lambda_0$.  The multiple minima for the $\EP$ ansatz are illustrated in
Figure~\ref{fig:energy}.  

\begin{figure}[htpb]
  \centering
\includegraphics[width=\columnwidth]{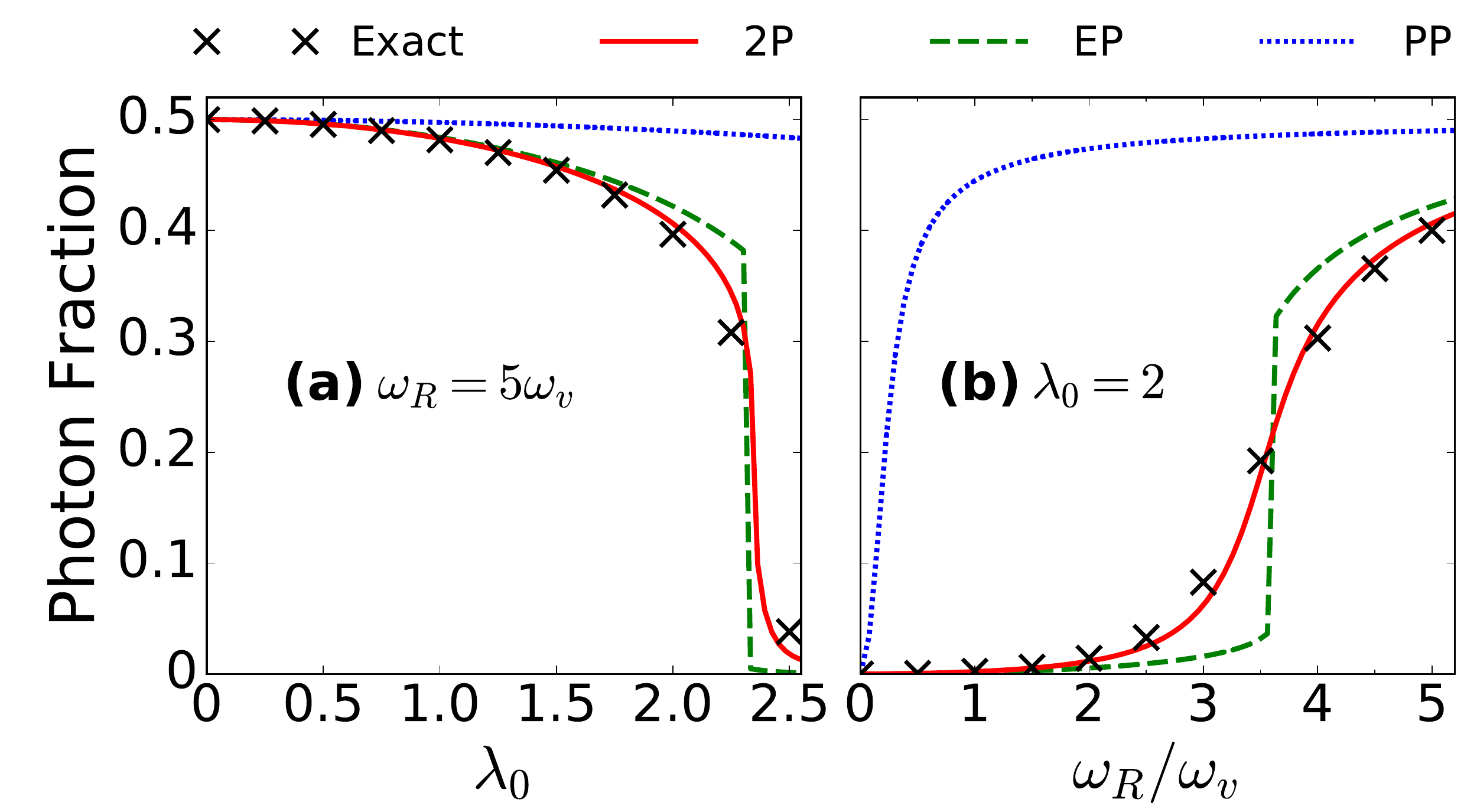}
  \caption{\new{Photon content of the lower polariton vs $\lambda_0$ and $\omega_R$ for $N=20$ molecules.  As in Fig.~\ref{fig:ELP}, the exact solution is shown as points, and the various approximations as lines.  
  For the EP and PP ans\"atze, the photon fraction is given by $|\alpha|^2$ from Eq.~(\ref{eq:EP}).
  For the 2P ansatz, the photon fraction from Eq.~(\ref{eq:TPS}) is 
  $\alpha_1^2+ 2\alpha_1\alpha_2 e^{-\lambda^2/2}+\alpha_2^2[1+(N-1)e^{-\lambda^2}]/N$. 
  In terms of the overall Wigner function in Eq.~(\ref{eq:composition}), the photon fraction refers to $\cos^2(\theta)$.}}
  \label{fig:composition}
\end{figure}

\new{As well as the sharp change in behavior visible from the ground state energy, a similar effect can be seen from the evolution of the photon fraction of the ground state, as shown in Fig.~\ref{fig:composition}.  In the limit of weak vibrational coupling, or strong Rabi splitting, we see the photon content of the lower polariton state approaches $0.5$, as expected for resonance between the photon and zero phonon line (i.e. $\omega=\omega_0)$.  When the vibrational coupling is strong or matter-light coupling is weak, the lower polariton instead becomes more excitonic, because the exciton energy including vibrational reorganization is $\omega_0 - \omega_v \lambda_0^2$ which, for the parameters we show, is smaller than $\omega$.}

\begin{figure}[ht]
  \includegraphics[width=3.2in]{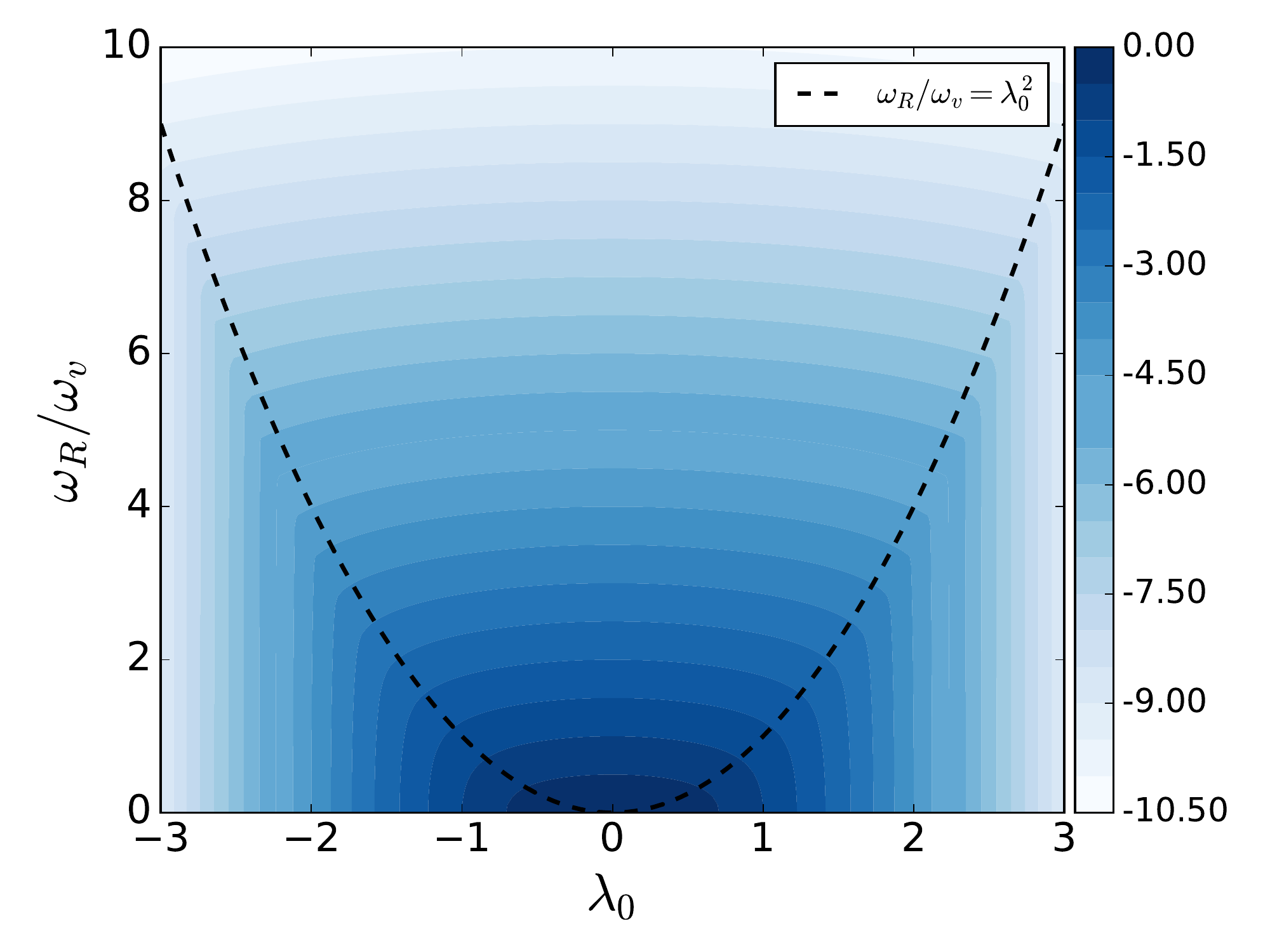}
  \caption{Optimized energy of $E_{\text{EP}}/\omega_v$ as a function of
    $\lambda_0$ and $\omega_R/\omega_v$.  The dashed line is the line
    $\omega_R = \omega_v \lambda_0^2$, and it can be seen that the locus of
    the discontinuity in energy closely follows this line at large $\lambda_0$,
    while at small $\lambda_0$, no discontinuity arises.  }
  \label{fig:energy-contour}
\end{figure}

To demonstrate that the change of characteristic behavior is indeed at
$\omega_R \simeq \omega_v \lambda_0^2$, Fig.~\ref{fig:energy-contour}
plots the energy landscape of $\EP$ vs the two relevant dimensionless
parameters $\lambda_0$ and $\omega_R/\omega_v$ (the energy of $\EP$ is
independent of $N$).  On this contour map, one can clearly see that the
locus of the discontinuity follows this line.  One may also note that a
sharp kink exists for $\lambda_0 \gtrsim 1.8$, otherwise the local minima
in $E_{\text{EP}}(\lambda)$ merge.

The failure of the $\EP$ state close to the kink is similar to the failure of the polaron
ansatz in the Rabi model~\cite{irish05,irish07}.  In the single molecule
case, it is known this can be remedied by an ansatz including a superposition
of two polaronic displacements~\cite{bera14}. The exact Wigner 
function in this regime clearly shows a bimodal distribution, indicating
such a  multi-polaron ansatz is required.  By
considering the structure of the vibrational state revealed by the exact
solution, we see that simultaneous displacement of multiple molecules is
not significant.  We therefore consider the following \textit{two polaron}
ansatz:
\begin{multline}
  \label{eq:TPS}
  \tPS
  =
  \ket{P} \otimes
  \frac{1}{N} \sum_i
  \left(
    \alpha_1
    +
    \alpha_2
    \mathcal{D}(\lambda, \bn_i) 
  \right) 
  \ket{0}_V
  \\ +
  \frac{1}{\sqrt{N}}\sum_i 
  \ket{X}_i \otimes
  \left(
    \beta_1
    +
    \beta_2
    \mathcal{D}(\lambda, \bn_i) 
  \right)
  \ket{0}_V.
\end{multline}
This ansatz is more challenging to numerically minimize, information on the procedures we use are given in the methods section.
  As seen from the energy in Fig.~\ref{fig:ELP}, the $\tPS$ ansatz
has no kink at $\omega_R \simeq \omega_v \lambda_0^2$ and so matches the
exact solution better than $\EP$.  
This is explained in
Fig.~\ref{fig:energy} which shows the energy landscape vs $\lambda$ for $\EP$ and
$\tPS$ near this point. One sees that for $\EP$ there are two minima, and so
a change of characteristic behavior occurs when these minima cross.  For
$\tPS$, the state is a superposition of the two minima of $\EP$, and so the
distinct local minima are washed out.

\begin{figure}[t]
 \includegraphics[width=3.2in]{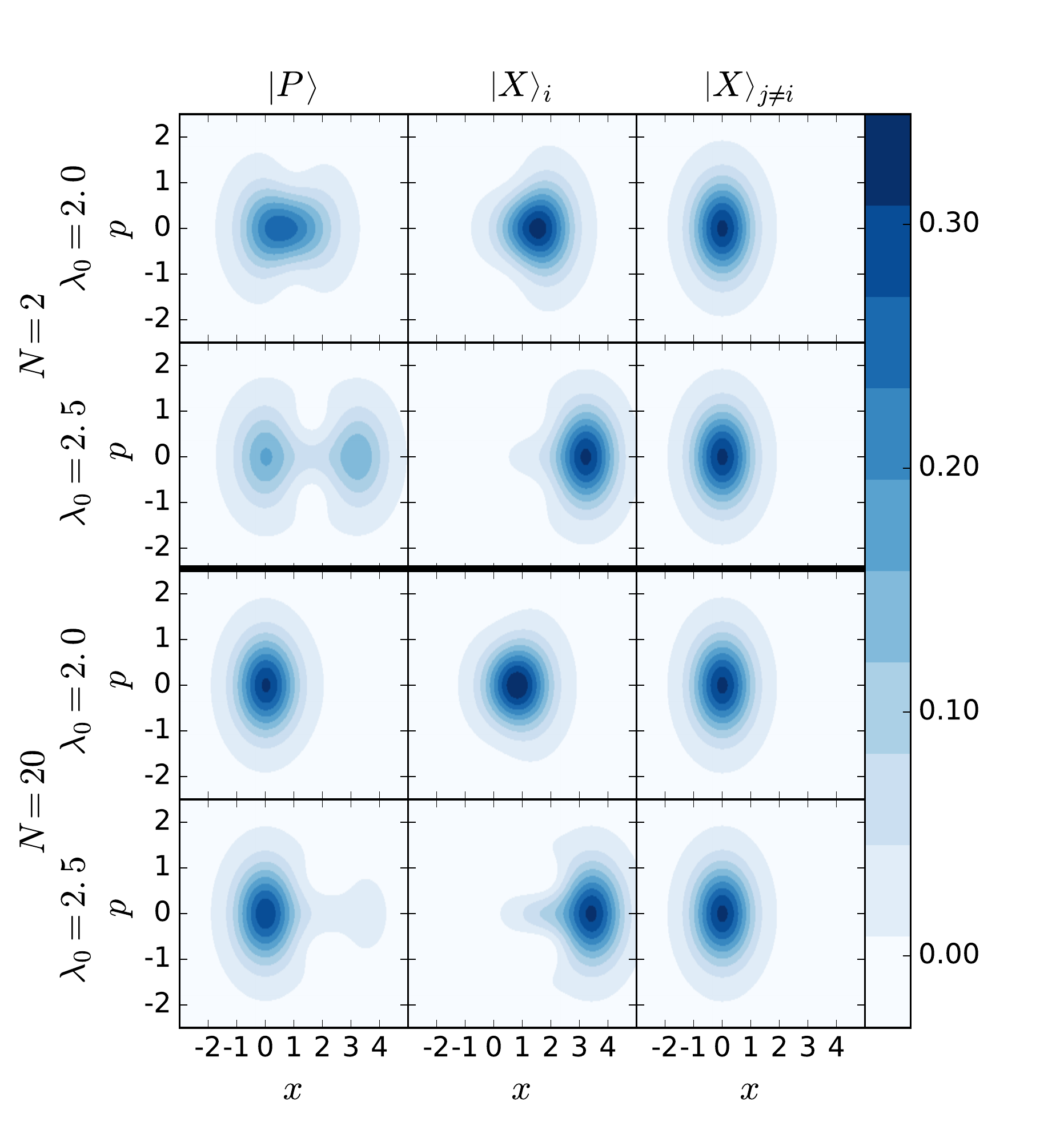}
 \caption{  For $\tPS$ ansatz, the conditional Wigner functions for $N=2$ (top) and $N=20$ (bottom) for the same parameters as in Figure~\ref{fig:wf}.
 \label{fig:wfa}
 }
 \end{figure}

Figure~\ref{fig:wfa} shows the conditional Wigner functions for the
same parameters as Fig.~\ref{fig:wf}, but calculated from the $\tPS$ ansatz. The $\tPS$ ansatz assumes the $\ket{X}_{j \neq i}$
vibrational state to be the vacuum so the right column is always the same
Gaussian.  At $\lambda_0=2.0$ (rows 1 and 3), as seen in the exact results,
the Wigner Function is close to the vibrational vacuum state almost
independent of the electronic state.  However, for $\lambda_0=2.5$ (rows 2 and
4), entanglement between the vibrational and electronic state is clearly
visible. The ansatz reproduces the numerical results, including the
suppression of entanglement with increasing $N$;  this suppression occurs only when $\omega_R>\lambda_0^2\omega_v$, otherwise the  variational parameter $\lambda$ saturates at $\lambda_0$ as $N$ increases. This demonstrates that
$\tPS$ captures not only the energy of the exact solution, but also
captures well the exact state.  i.e. Eq.~(\ref{eq:TPS}) can be
used to provide an accurate analytic description of the lower polariton state in
future work.

\subsection*{Absorption spectrum}

In the previous sections, we discussed the nature of the 
lower polariton state, with consequences on whether vibrational
decoupling occurs.  Here we show how the nature of this state 
is reflected by more routinely measured observables, specifically  
the absorption spectrum. 

In the literature, there are two strands of work regarding how the strong
coupling spectrum of vibrationally dressed excitons should be considered.
In many  works fitting experimental spectra~\cite{Kena-Cohen08, Plumhof14, Daskalakis14, Michetti15, dietrich16}, the vibrational sidebands of the molecule
are treated as if they corresponded to separate species of emitters.
i.e., one uses a ``coupled oscillator'' model, which describes 
a photon mode coupled to multiple excitonic resonances,
one for each vibrational sideband.  One can write such an approach
both as a quantum Hopfield model, or as a classical linear optics
calculation, with a material dielectric constant including all
transitions from the ground state to vibrational sidebands, 
i.e. it assumes vibrational
dressing of the electronic ground state is irrelevant.  Given
the results seen in Fig.~\ref{fig:wf} that indicate vibrational displacement
for unexcited molecules, this result is concerning.  An alternate
approach is to consider diagonalising the full HTC model in the one
excitation sector.  Approximate methods for finding the spectra at
intermediate $N$ have recently been discussed~\cite{Herrera2016a,Herrera2016}.
Here we  show how exact numerics at intermediate $N$ allows us to
demonstrate the validity of simple approximate expressions valid in the
thermodynamic limit, but show that more complicated results occur
at small $N$.

\begin{figure}
  \centering
  \includegraphics[width=\columnwidth]{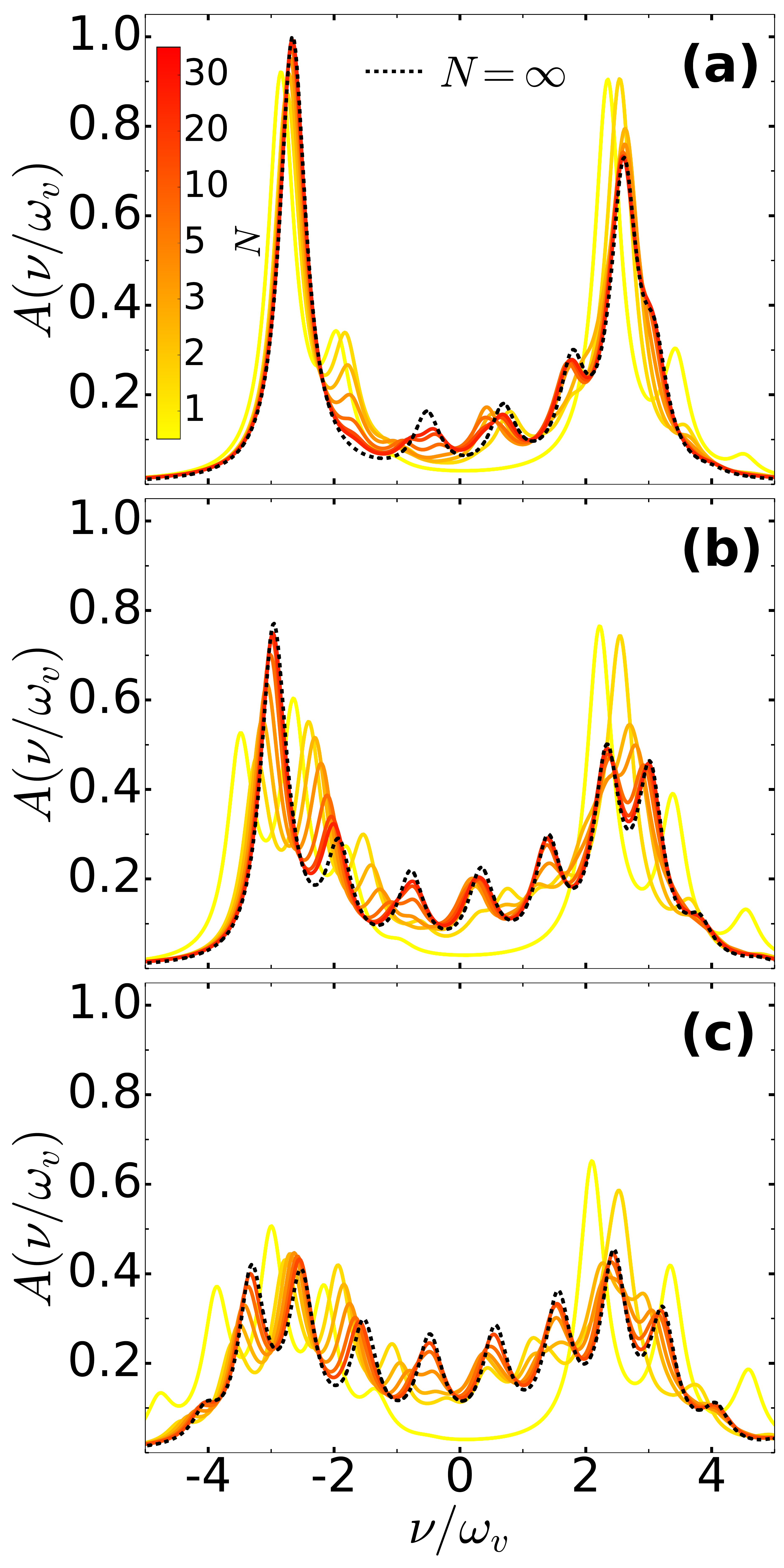}
  \caption{
  The evolution of absorption spectrum $A(\nu)$ with number of molecules $N$.
We compare the exact solutions (solid) for varying
    number of molecules $N$, against the analytic Green's  function (dotted) valid
    at $N \to \infty$
     at $\omega_R=2.5\omega_v$ and $\kappa=\gamma=0.5\omega_v$ 
     for three values of $\lambda_0=1, 1.5, 2$ in panels (a), (b), and (c) respectively.
    All curves are plotted for $M\geq5$ and $M_x\geq20$
    as discussed in the text.}
  \label{fig:abspl}
\end{figure}

Figure~\ref{fig:abspl} shows how the absorption spectrum\cite{Ciuti2006,Cwik16} 
$A(\nu) = -\kappa\left[\text{Im}[D^R(\nu)] +\kappa | D^R(\nu)|^2/2\right]$
evolves with number of molecules $N$ for three different values of $\lambda_0$. Here, $\kappa$ is the line-width of cavity (with both mirrors assumed identical) and $D^R(\nu)$ is the photon Green's function.
  We 
show numerical results for intermediate $N$, and  compare them to
a calculation based on the coupled oscillator model.  The analytic
expression is given by writing
$[D^R(\nu)]^{-1} = \nu + i \kappa/2 - \omega - \Sigma_X(\nu)$, where
$\Sigma_X(\nu)$ is an excitonic self energy.  In the large $N$ limit it is
expected~\cite{eastham01,cwik14,Cwik16} that this self energy can be
calculated considering each molecule separately, leading to a simple form,
\begin{equation}
  \label{eq:sigma}
  \Sigma_X(\nu) = \sum_m \frac{\omega_R^2 \left|f_m\right|^2%
  }{\nu + i\gamma/2 - (\omega_0 -\lambda_0^2\omega_v+ m \omega_v)},
\end{equation}
where $f_m=\langle m| \mathcal{D}(\lambda_0,\hat b^\dagger)|0\rangle$ is
the overlap between the $m$th excited vibrational state and the displaced
ground state.  This expression is clearly independent of $N$ (as the
$1/\sqrt{N}$ scaling of the matter-light coupling in Eq.~(\ref{eq:dh})
cancels with the sum over molecules).  

As can be seen from Fig.~\ref{fig:abspl}, remarkably the analytic expression  matches
the exact results at $N\to \infty$, validating that such an
expression applies in this limit, but clearly differs at small $N$.  

To ensure convergence of the numerical results we include a much larger number of 
vibrational states, $M_x$, for the electronically excited molecule
compared to the cutoff for
electronically unexcited molecules at $M$. Both of these cutoffs
vary with $N$.
In particular, for small $N$, we find that more vibrational states must be kept.  In all
cases shown, convergence with increasing $M$ and $M_x$ was achieved, the smallest
value used (for $N=30$) was $M=5$ and $M_x=20$.

The evolution of the spectral weight with increasing $N$ shows some
interesting features. At small $\lambda_0$ when $N=1$ there is only a single vibrational replica
close to the large lower and upper polariton peaks.  As $N$ initially increases, other
peaks grow, and then at very large $N$, several of the initially visible
peaks vanish.  The vanishing of peaks at large $N$, and hence the match to the coupled oscillator model, can be
straightforwardly explained: these peaks involve transitions to
vibrationally excited states in the electronic ground state manifold. 
Such states are absent from Eq.~(\ref{eq:sigma}), as can be seen from the
fact that this expression involves only transitions between the single
state $\ket{0}$ in the ground state manifold, and a sequence of states in
the excited manifold.
For small $N$, transitions to such states have an effect, hence the small
$N$ curves in Fig.~\ref{fig:abspl} do not match Eq.~(\ref{eq:sigma}).
However, at large $N$, these transitions are suppressed.  Physically, the
Green's function corresponds to an amplitude to evolve from a state $\ket{\alpha}=a^\dagger |0\rangle \otimes |0\rangle_V$
back to $\ket{\alpha}$, i.e.\ to return back to a state without any
vibrational excitations.  Thus, any transition that excites a vibrational
mode must be compensated by a subsequent transition (or transitions)
returning a molecule to its vibrational ground state.  However the
probability that a single molecule is excited multiple times falls as a
power of $N$.  Hence, in the large $N$ limit, each molecule is excited
only once, recovering Eq.~(\ref{eq:sigma}). Similar behavior is seen as $\lambda_0$ increases. The competition between polaron and polariton physics makes all of the peaks similar in height.

\section*{Methods}

\subsection*{Exact numerical solution}
\label{sec:exact-numerics}


Here we focus on results that require only
the $N_{\text{ex}}=1$ sector, but similar approaches allow extension to
higher sectors.  Within the $N_{\text{ex}}=1$ sector, possible states 
are spanned by the product between states
\begin{align}
  \label{eq:xp}
  \ket{X}_i &= \ket{0}_P\ket{\da\ldots\ua_i\ldots\da}, \\
  \ket{P} &= \ket{1}_P\ket{\da\ldots\da},
\end{align}
and vibrational states $|\{n_i\}\rangle_V$ (where $\{ n_{i}\}$ count the
vibrational quanta on each molecule).

Our numerical approach makes use of the fact that both the minimum
energy state and the absorption spectrum of the system involve only
states with full permutation symmetry between molecules.
Such states, $\ket{\Psi}$ satisfy
\begin{multline}
  \bra{\Psi} 
  \left(\ket{X}_j \otimes
  \ket{n_{1}\ldots  n_{j} \ldots  n_{k} \ldots  n_{N}}_V\right)
  \\=
  \bra{\Psi}\left(\ket{X}_k
  \otimes \ket{n_{1}\ldots  n_{k} \ldots  n_{j} \ldots  n_{N}}_V\right),
\end{multline}
for the excited molecule, along with symmetry under any permutation of the
vibrational states of the $N-1$ unexcited molecules.  Similarly the
coefficients $\langle \Psi| (\ket{P} \otimes |\{ n_i \} \rangle_V) $ have
symmetry under permutations of all $n_i$.  Among the symmetrically related
coefficients, we can choose a single unique pattern $\{n_i\}$ and use this
to represent all equivalent states.  Truncating the number of
vibrational quanta on a site to $n_i \leq M$, the symmetry reduces the
problem size from a matrix of size $\mathcal{O}([M+1]^N)$ to
$\mathcal{O}(N^M)$; i.e.\ from exponential in system size to polynomial. 
When $N$ is large,
the basis size can still be kept to a manageable size without significantly compromising accuracy by 
using different number of vibrational states for 
electronically unexcited and excited molecules, $M$ and $M_x$ with $M_x > M$.
Similar techniques have recently been used to study non-equilibrium steady states of related models~\cite{Kirton2017}.
Full details of how to index these states is provided in the supplementary material~\footnote{Supplementary material, containing details of the
polynomial-scaling algorithm.}.

\subsection*{Variational expressions}
\label{sec:vari-expr}

Minimization of the variational ans\"atze are performed numerically.
For the
$\EP$ ansatz the expression for the energy turns out to be 
simple.
 For the $\tPS$, the result
is more complicated.
We first consider the normalization of the $\tPS$ state.
For the photonic part, one must be careful to include cross
terms  between different contributions in the sum, whereas for the
excitonic part, entanglement with the electronic state means
no cross terms exist.  
We will denote the cross term overlap as
\begin{math} \mathcal{O}\equiv \langle 0 | 
  e^{\lambda (\bd_i  - \bn_i)}
  | 0 \rangle = e^{-\lambda^2/2}
\end{math}, in terms of which we find:
\begin{multline*}
  \braket{\text{2P}|{\text{2P}}}
  \\=
  \frac{1}{N^2} \sum_{ij} \biggl[
    \alpha_1^2
    + 2 \alpha_1 \alpha_2 \mathcal{O}
   +\alpha_1^2 \left( \delta_{ij}  + (1-\delta_{ij})  \mathcal{O}^2 \right)
  \biggr] 
  \\+
 \beta_1 ^2 + 2 \beta_1 \beta_2 \mathcal{O} + \beta_2^2.
 \end{multline*}
This normalization condition 
can be
written as a quadratic form in terms of the
linear coefficients $\phi^T=(\alpha_1, \alpha_2, \beta_1, \beta_2)$. This means
we may write , $ \phi^T \mbf{N} \phi = 1$, where
the matrix $\mbf{N}$ in block form is:
 \begin{align*}
  \mbf{N}
  &=
  \begin{pmatrix}
    \mbf{N}^\alpha & 0 \\ 0 & \mbf{N}^\beta
  \end{pmatrix}, \quad \text{where}
  \\
  \mbf{N}^\alpha &= 
  \begin{pmatrix}
   1
    &
    \mathcal{O}   \\
   \mathcal{O}   &
    \left[1 + \left(N-1 \right)  \mathcal{O}^2\right]/N
  \end{pmatrix}, \\
  \mbf{N}^\beta &=
  \begin{pmatrix}
    1 &\mathcal{O}\\\mathcal{O}& 1
  \end{pmatrix}.
\end{align*}

To calculate the energy, we split it in two parts.  The non-vibrational
terms (i.e. those from the Tavis-Cummings Hamiltonian) are relatively simple,
as the photonic and electronic energies just follow the structure above, and
the matter-light coupling term has the same matrix structure as the photon part, thus:
\begin{displaymath}
  E_{\text{TC}}
  = \phi^T \mbf{M}_{\text{TC}} \phi, \quad
  \mbf{M}_{\text{TC}} = 
  \begin{pmatrix}
    \omega \mbf{N}^\alpha & \omega_R \mbf{N}^\alpha \\
    \omega_R \mbf{N}^\alpha & \omega_0 \mbf{N}^\beta
  \end{pmatrix}.
\end{displaymath}
We must then add to this the vibrational part of the energy.  This has the
form:
\begin{multline}
  \frac{E_{v}}{\omega_v}=
  \frac{\alpha_2^2 \lambda^2}{N}
  +
  \beta_2^2 (\lambda^2 - 2 \lambda \lambda_0)
  -2 \beta_1 \beta_2 \mathcal{O}\lambda\lambda_0.
\end{multline}
Combining both parts we write $ E_{\text{2P}} =
\phi^T \mbf{M} \phi$ where:
\begin{equation}
  \mbf{M}
  =
  \begin{pmatrix}
    \omega \mbf{N}^\alpha + \omega_v \mbf{M}_{v}^{\alpha}
    & 
    \omega_R \mbf{N}^\alpha 
    \\
    \omega_R \mbf{N}^\alpha 
    & 
    \omega_0 \mbf{N}^\beta + \omega_v \mbf{M}_{v}^{\beta}
  \end{pmatrix},
\end{equation} 
with the extra block matrices: 
\begin{align*}
  \mbf{M}_{v}^{\beta} &= 
  \begin{pmatrix}
   0 & 
    -\lambda\lambda_0\mathcal{O}\\
    -\lambda\lambda_0\mathcal{O}&
    \lambda^2 - 2 \lambda \lambda_0 
  \end{pmatrix}
  , \\
  \mbf{M}_{v}^{\alpha} &= 
  \begin{pmatrix}
     0 &0\\
     0& \lambda^2/N
  \end{pmatrix}.
\end{align*}
In terms of the above expressions, the energy corresponds to minimization of
$\phi^T \mbf{M} \phi$ subject to the constraint $\phi^T \mbf{N} \phi=1$. Thus the ground
state energy corresponds to solving the generalized
eigenvalue problem $\mbf{M} \phi = \mu(\lambda) \mbf{N} \phi$, and
then minimizing over $\lambda$.  We perform this minimization using a
conjugate gradient approach from the SciPy library~\cite{scipy}.

It is instructive to note that at $N\to \infty$, there remains a regular form
of all matrices above.  This has the immediate consequence that
at $N \to \infty$, we get a finite asymptote for the coefficients
$\alpha_{1,2}, \beta_{1,2}$.

\subsection*{Calculating retarded Green's functions}
\label{sec:calc-retard-greens}

The reflection, absorption and emission spectra of the coupled
cavity-molecule system can all be found in terms of the retarded photon
Green's function~\cite{Ciuti2006,Cwik16,Herrera2016}, 
\begin{equation}
 D^R(t) = -i
\Theta(t) \bra{0} [ \an(t), \ad(0) ] \ket{0}.
\end{equation}
In
particular, the absorption spectrum $A(\nu)$ is given by 
\begin{equation}
 A(\nu) = -
\kappa_{L} \left[ 2 \text{Im}[D^R(\nu)] + (\kappa_{L} +
  \kappa_{R}) |D^R(\nu)|^2\right], 
\end{equation}
where $\kappa_{L,R}$
denote the losses through ``left'' and ``right'' mirrors of the cavity.  
In our plots, we assume a balanced cavity with $\kappa_L=\kappa_R$.
The retarded Green's function
can be found exactly by time evolution within the permutation symmetric
states: i.e.\ one needs to find $\langle \alpha | e^{-i \hat H t}
|\alpha\rangle$ where $\ket{\alpha}$ is a state with a single photon and
all molecules in their electronic and vibrational ground state.

In evaluating this, it is useful to reduce the basis size by exploiting 
a vibrationally displaced basis set that lowers the vibrational cutoff $M$. 
In such a case, the vibrational part of the vacuum state becomes a multidimensional coherent state.

To calculate a spectrum with finite linewidths, we incorporate decay rates for
both the cavity, $\kappa$ and electronic excitations, $\gamma$, via
non-Hermitian terms in the Hamiltonian. i.e. we write
 $H_{decay}=-i(\kappa\ket{P}\bra{P} + \gamma \ket{X}\bra{X})/2$ in the
Hamiltonian for approximating the cavity and exciton decay rates, $\kappa$
and $\gamma$. Here, $\ket{P}\bra{P}$ and $\ket{X}\bra{X}$ are respectively
projectors onto the subspace with a photon or exciton present. 

\section*{Summary}
\label{sec:disc-concl} 
In summary, we have presented exact solutions of the
Holstein--Tavis--Cummings model at intermediate $N$, and used this to identify  highly
accurate analytic approaches which capture the behavior seen and can be extrapolated to $N \to \infty$. 

Regarding the nature of the lower polariton state, we find
the behavior divides into two distinct classes: suppression of vibrational
entanglement, and the consequent physics discussed in Ref.~\citen{Herrera15}
when $\omega_R \gg \omega_v \lambda_0^2$, and a sharp crossover to
entangled behavior when these scales become comparable. By considering
the absorption spectrum, we have also
shown how these entangled polaritonic states are nonetheless consistent
with the previously used~\cite{cwik14,Cwik16} form of absorption spectrum
recovered in the $N\to \infty$ limit.  

The method of exact solution we used
applies to a variety of generalizations of the Holstein--Tavis--Cummings
model in the absence of disorder.  For example, in future work, such exact
solutions can be used to address the role of interactions or to
go beyond the rotating wave approximation.  The analytic ansatz presented
provides an explicit description of the behavior of the model in the
thermodynamic limit, and can provide a complementary tool to the methods
proposed in Ref.~\citen{galego16}. In the future, it may also provide a basis to
develop the polaron master
equation~\cite{McCutcheon10,McCutcheon11,Roy12,Pollock13} for $N$-molecule
systems.

\begin{acknowledgement}
  JK and MAZ acknowledges financial support from EPSRC program ``Hybrid
  Polaritonics'' (EP/M025330/1).  PGK acknowledges support from EPSRC
  (EP/M010910/1).  We are grateful to F.~Herrera, B.~W.~Lovett, and
  M.~C.~Gather for helpful comments on a previous version of the manuscript.
\end{acknowledgement}

\providecommand{\latin}[1]{#1}
\makeatletter
\providecommand{\doi}
  {\begingroup\let\do\@makeother\dospecials
  \catcode`\{=1 \catcode`\}=2 \doi@aux}
\providecommand{\doi@aux}[1]{\endgroup\texttt{#1}}
\makeatother
\providecommand*\mcitethebibliography{\thebibliography}
\csname @ifundefined\endcsname{endmcitethebibliography}
  {\let\endmcitethebibliography\endthebibliography}{}

\renewcommand{\figurename}{Supplemental Figure}
\renewcommand{\theequation}{S\arabic{equation}}
\setcounter{equation}{0}
\setcounter{figure}{0}
\setcounter{section}{0}

\clearpage
\onecolumn

\begin{center}
\textbf{\large Supplemental Material: Exact states and spectra of vibrationally dressed polaritons}

\vspace{0.4cm}

{M.\ Ahsan Zeb, Peter G.\ Kirton and Jonathan Keeling} \\
\textit{SUPA, School of Physics and Astronomy, University of St Andrews, St Andrews, KY16 9SS, United Kingdom}\\
(Dated: \today)

\end{center}

\begin{multicols}{2}

This supplemental material provides further details of the
polynomial-scaling algorithm used to exactly solve the
Holstein--Tavis--Cummings model and get the conditional reduced vibrational density matrices for the three conditions discussed in the paper.
Section~\ref{sec:basis-set} describes the set of basis
states, section~\ref{sec:hamilt-matr-elem} the form of the Hamiltonian
matrix elements in this basis, and section~\ref{sec:cond-dens-matr}
discusses how to extract reduced conditional density matrices from the
wavefunction in the given representation. 

\section{Basis set}
\label{sec:basis-set}

The basis states can be divided into two distinct subspaces, corresponding
to whether  one has states $\ket{X}_i$ or $\ket{P}$.  

For the states corresponding to $\ket{X}_i$ (which we call the excitonic
subspace), one molecule is excited, and $N-1$ are unexcited.  The vibrational
state of the excited molecule is represented explicitly by the
occupation $m^\ast \in [0,M]$.  For the $N-1$ unexcited molecules, the vibrational
state must be permutationally symmetric, thus we may pick a representative
state $\{m_{j \neq i}\}$ to stand for the superposition of all states related
by permutation symmetry, $\ket{\symset_{N-1}{\{m_j\}}}$.  To uniquely specify the representative state, we
choose the state where the occupations are in ascending order.  For
example, with $N-1=4$, we may denote $\ket{\symset_{4}\{0112\}}$ as
representing the linear superposition of all permutations of this pattern of
occupations:
\begin{displaymath}
\ket{\symset_{4}\{0112\}} \equiv 
\frac{
\begin{aligned}
\bigl(&\ket{0112} + \ket{1012} + \ket{1102} + \ket{1120} \\ 
+& \ket{0211}  + \ket{2011}  + \ket{2101}  + \ket{2110} \\ 
+& \ket{0121}+ \ket{1021}+ \ket{1201}+ \ket{1210} \bigr)
\end{aligned}
}{\sqrt{\cntset_4(\{0112\})}}
\end{displaymath}
where $\cntset_N(\{m_j\})$ counts the number of distinct permutations, i.e.
$\cntset_4(\{0112\})=12$ for this  example.

The complete vibrational state, where molecule $i$ is excited, can then be
written as $\ket{m^\ast_i,\symset_{N-1}\{m_{j \neq i}\}}_V$.  Since states
with other molecules excited are related by permutation symmetry, 
the most general state with an excited molecule can be written as:
\begin{multline}
  \ket{X}\otimes\ket{m^\ast,\symset_{N-1}\{m\}}_V
  \\\equiv 
  \frac{1}{\sqrt{N}}\sum_{i=1}^N\ket{X}_i\otimes\ket{m^\ast,\symset_{N-1}\{m_{j\ne i}\}}_V.
\end{multline}
If the occupation of the vibrational modes is restricted to $m \leq M$, then the number
of distinct coefficients for an excited molecule is $(M+1) \cdot
{}^{M+N-1}C_{M}$, where the combinatoric factor is the number of distinct
sets $\{m\}$.

For the states with a photon present (i.e. the photonic subspace), no molecule
is electronically excited, so there is full permutation symmetry.  These
states can thus be written immediately as
\begin{gather}
  \ket{P}\otimes\ket{\symset_{N}\{m\}}_V.
\end{gather}
As the set of permutations now refers to $N$ molecules, the total
number of such states is just  ${}^{M+N}C_{M}$.

\section{Matrix elements of Hamiltonian }
\label{sec:hamilt-matr-elem}

\subsubsection{Diagonal Terms}
The terms $\ad \an$, $ \sum_i \hat \sigma^+_i \hat \sigma^-_i$ are diagonal in
the above basis, and have the value $0(1)$ or $1(0)$ respectively in the
photon(exciton) subspace.  The term $\sum_i \hat b^\dagger_i \hat b^{}_i$ is
also diagonal, and is given by $m^\ast+\sum_i m_i$ in the exciton block, and $\sum_i
m_i$ in the photon block.

\subsubsection{Vibrational Coupling}

The term coupling the electronic and vibrational states,
\begin{gather} 
H_{\lambda}=\sum_i\sigma^+_i \hat \sigma^-_i ( \hat b^\dagger_i + \hat b^{}_i ), \nonumber
\end{gather}
is not diagonal, but is straightforward. It vanishes when acting on
states involving $\ket{P}$.  For states involving $\ket{X}_i$, it is
diagonal in the vibrational state of the other $N-1$ molecules, and
so takes the form:
\begin{multline} 
  \Bigl( \bra{X}\otimes\bra{m^{\ast \prime},\symset_{N-1}\{m\}}_V\Bigr)
  H_\lambda
  \Bigl( \ket{X}\otimes\ket{m^\ast,\symset_{N-1}\{m\}}_V \Bigr)
  \\= 
  \sqrt{m^{\ast\prime}} \delta_{m^\ast-1,m^{\ast\prime}}+
  \sqrt{m^\ast}\delta_{m^\ast,m^{\ast\prime}-1}\nonumber
\end{multline}
where $m^\ast$ and $m^{\ast\prime}$ are vibrational occupations of the excited
molecule.

\subsubsection{Matter-Light Coupling}

The only complicated term is the matter-light coupling:
\begin{displaymath}
  H_{R}= \sum_i\left(\hat \sigma^+_i \hat a^{} + \hat \sigma^-_i \hat
    a^\dagger\right),
\end{displaymath}
as this mixes the two sectors, which are labeled differently. 

We focus on the photon emission term, as the other term follows by
conjugation.  We need to find:
\begin{multline}
  \Bigl( \bra{P}\otimes\bra{\symset_{N}\{m^\prime\}}_V \Bigr)
  \sum_i \hat \sigma^-_i \hat a^\dagger   
  \Bigl( \ket{X}\otimes\ket{m^\ast,\symset_{N-1}\{m\}}_V \Bigr)
  \\\equiv
  O_{\{m^\prime\}, m^\ast, \{m\}} 
\end{multline}
In order for this not to vanish, the set of final occupations $\{m^\prime\}$ must
have an element equal to $m^\ast$ and the rest of the set equal to $\{m\}$.
If this is true, then we have:
\begin{displaymath}
  O_{\{m^\prime\}, m^\ast, \{m\}} 
  = 
  \frac{N \cntset_{N-1}(\{m\}) }{\sqrt{N \cntset_{N-1}(\{m\}) } \sqrt{\cntset_N(\{m^\prime\})}}
\end{displaymath}
where $\cntset_{N-1}(\{m\})$ is the count of distinct permutations as defined
above.  The factors $\sqrt{N \cntset_{N-1}(\{m\})}$ and
$\sqrt{\cntset_N(\{m^\prime\})}$ in the denominator come from the normalization
of $\ket{m^\ast,\symset_{N-1}\{m\}}_V$ and $\ket{\symset_{N}\{m^\prime\}}_V$.
The factors in the numerator come from the distinct ways in which
the overlap may occur --- a choice of $N$ molecules that may be excited, and
$\cntset_{N-1}(\{m\})$ ways of arranging the unexcited molecules in the overlap.

The conditions given above means that by taking a single element $m^\ast$ out
of $\{m^\prime\}$, the rest should become equal to $\{m\}$.  Since
coefficients are indexed over these sets, we can consider this as requiring a
mapping $\maptoN$ from the pair of
integers $(m^\ast, \indset_{N-1}(\{ m \}))$, to the integer
$\indset_N(\{m^\prime\})$, where $\indset_N(\{ m \})$ is the index of an ordered
set of $N$ occupations.  We compute this mapping once and store for later use.

\section{Conditional Density Matrices}
\label{sec:cond-dens-matr}

In the paper, we plot the conditional Wigner functions of the vibrational
state of a single molecule.  These correspond to the reduced density matrices
subject to the three possible conditions, denoted by $\ket{P}, \ket{X}_i$ and
$\ket{X}_{j\neq i}$ in the paper.  By using the basis states described above,
we can numerically diagonalize to find the ground state wavefunction
$\ket{\psi}$ in the basis of states described above.  This section discusses how
to extract the corresponding reduced conditional density matrices from such a
representation of the state.
 
\paragraph{Photonic sector $\ket{P}$}
\label{sec:photonic-sector}

We calculate the conditional reduced density matrix $\rho^{\ket{P}}$ which corresponds to
restricting to the subspace
involving $\ket{P}$.  We then need to trace out the vibrational 
state of $N-1$ of
the molecules.  To do this, we can re-use the map $\mathcal{M}_{N-1
  \mapsto N}$.  This mapping finds which set of indices for the  $N-1$ 
molecules we trace out combine with the state $m$ of the molecule of interest
to find a given index for the full $N$  molecule problem.  
To find the element $\rho^{\ket{P}}_{m,m^\prime}$, we need to find all pairs
of states of the $N$ molecule problem which are reduced to the same $N-1$
molecule state when $m,m'$ are taken out. For example, if we denote
$j^{}_N = \indset_N(\{m\})$ as an index of a state of $N$ molecules, 
we need to find states such that
\begin{equation}
  \label{eq:define_N_map}
  \begin{gathered}
  k^{}_N    = \maptoN(m,      j_{N-1}), \\
  k^\prime_N = \maptoN(m^\prime,j_{N-1}).
  \end{gathered}
\end{equation}
This is so that  we can trace over $j_{N-1}$, describing the state of
the other molecules.  Using these labels we can then write
\begin{equation}
  \label{eq:reduce_P}
  \rho^{\ket{P}}_{m,m'} = \sum_{j_{N-1}=1}^{\mathcal{N}_{N-1}}
  \frac{\psi^\ast_{P,k_N} \psi^{}_{P,k^\prime_N} \cntset_{N-1}(j_{N-1})
  }{\sqrt{\cntset_{N}(k_{N}) \cntset_{N}(k^\prime_{N})}} 
\end{equation}
where the summation is over all $\mathcal{N}_{N-1} \equiv {}^{N-1+M}C_M$
vibrational states of $N-1$ molecules.  Note that the indices $k_N,k^\prime_N$
appearing here are related to $j_N$ by Eq.~(\ref{eq:define_N_map}).  As
defined above, $\psi_{P,k_N}$ is the coefficient of the state with index
$k_N$ in the $\ket{P}$ subspace, and we have used the number of distinct
permutations $\cntset_N(k_N)$ as shorthand for $\cntset_N(\{m\})$ where
$k_N=\indset_N(\{m\})$.  The factors in the denominator come from the
normalization of the basis states, and the factor in the numerator comes
from counting the number of terms contributing to the trace.

\paragraph{Excited molecule, $\ket{X}_i$}

For $\rho^{\ket{X}_i}$, the conditional reduced density matrix is straightforward,
as the vibrational state of the excited molecule is represented explicitly.
We need only to trace over the other $N-1$ molecules.  Using
$j_{N-1}$ as the index of these states we find
\begin{equation}
  \label{eq:reduce_Xi}
  \rho^{\ket{X}_i}_{m,m'} = 
  \frac{1}{N}
  \sum_{j_{N-1}=1}^{\mathcal{N}_{N-1}}
  \psi^\ast_{X,m,j_{N-1}}  \psi^{}_{X,m^\prime,j_{N-1}}  
\end{equation}
Here $\psi_{X,m,j_{N-1}}$ is the coefficient of the state in with index
$(m,j_{N-1})$ in the $\ket{X}$ subspace, and the factor of $N$ appears
from the normalization of basis states.

\paragraph{Other molecule excited, $\ket{X}_{j \neq i}$}

To calculate $\rho^{\ket{X}_{j \neq i}}$, the conditional reduced density
matrix corresponding a different molecule being electronically excited,
we must trace over the excited molecule, and over $N-2$ of the $N-1$
unexcited molecules.  This is similar to the case for
$\ket{P}$, but this time we must use the mapping
$\maptoNa$, from the indexing of distinct patterns of $N-2$ 
molecules and one explicit index, to the indexing of $N-1$ molecules.

\begin{equation}
  \label{eq:define_N1_map}
  \begin{gathered}
  k^{}_{N-1}    = \maptoNa(m,      j_{N-2}), \\
  k^\prime_{N-1} = \maptoNa(m^\prime,j_{N-2}).
  \end{gathered}
\end{equation}
Then, using similar notation to Eq.~(\ref{eq:reduce_P}) we may write:
\begin{multline}
  \label{eq:reduce_Xjnoti}
  \rho^{\ket{X}_{j \neq i}}_{m,m'} = 
  \sum_{j_{N-2}=1}^{\mathcal{N}_{N-2}}
  \sum_{m^{\prime\prime}=0}^{M}
  \psi^\ast_{X,m^{\prime\prime}, k_{N-1}}
  \psi^\ast_{X,m^{\prime\prime}, k^\prime_{N-1}}
  \\\times
  \frac{N-1}{N}
  \frac{\cntset_{N-2}(j_{N-2})}{%
    \sqrt{\cntset_{N-1}(k_{N-1}) \cntset_{N-1}(k_{N-1}^\prime)}}.
\end{multline}

\end{multicols}
\end{document}